\documentclass{eas}
\usepackage{graphicx}
\begin{document}

\title{The Disk-Halo Connection and Where Has All The Gas Gone?} 
\author{Joel N. Bregman}\address{Department of Astronomy, University of Michigan}
\begin{abstract}
The wealth of data in the past decades, and especially in the past 15 years 
has transformed our picture of the gas around the Milky Way and other 
spiral galaxies.  There is good evidence for extraplanar gas that is a few 
kpc in height and is seen in all gaseous phases: neutral; warm atomic; and 
hot, X-ray emitting gas.  This medium is seen not only around the Milky Way, 
but other spiral galaxies and it is related to the star formation rate, 
so it is likely produced by the activity in the disk through a galactic fountain.  
More extended examples of halo gas are seen, such as the HVC around the 
Milky Way and around M31.  This gas is typically 10-20 kpc from the galaxy 
and is not seen beyond 50 kpc.  This gas is most likely being accreted. 
A hot dilute halo (10$^6$ K) is present with a similar size, although 
its size is poorly determined.\\
An ongoing controversy surrounds the relative amounts of outflow from
the disk and accretion onto galaxies such as the Milky Way.  There is good
evidence for accretion of cold material onto the Milky Way and other galaxies, 
but it is not clear if there is enough to modify the gas content and star 
formation properties in the disk.  The reservoir of accretion material is as
yet unidentified.  Some of these findings may be related to the issue that 
galaxies are baryon-poor:  
their baryon to dark matter ratio is well below the cosmological
value.  The absence of baryons may be due to extremely violent outflow
events in the early stages of galaxy formation.  We may be able to
understand this stage of galaxy evolution by applying our deepening
understanding of our local disk-halo environment.
\end{abstract}
\maketitle
\runningtitle{Overview of the Disk-Halo Connection}
\section{Introduction}

About 50 years ago, the concept of gas above the disk of the Milky Way
was suggested by Spitzer (1956), based on properties of optical absorption
line systems.  In a remarkable paper, he discussed the nature of gas that
might occupy the volume above the disk, and its possible evolution.  At
that time, there were few observational tools to probe these suggestions,
but that has changed enormously in the following decades.  In the 1960's,
the discovery of the high velocity clouds of neutral hydrogen was a
dramatic observational discovery and nearly every theoretical explanation
placed this material above the disk (Oort 1970).  Also, the amount of mass
in the HVC could be substantial, depending on the distance to these
clouds, which became one of the most difficult problems to resolve.

The next great observational insights occurred in the mid 1970s with the
opening of two new wavebands, the X-ray and the rocket ultraviolet
regions.  The X-ray observations showed that gas of ${\sim}${}10$^{{\rm 6}}$ K is present in
the Milky Way and that the Sun is located in a hot bubble.  The presence
of large million degree gas regions was quickly interpreted as the natural
consequence of supernovae and led to a fundamentally new vision for our
interstellar medium (on a personal note, as a graduate student, I first heard
about this in a colloquium given by Don Cox, which had an important
effect on my career).  Our ISM no longer resembled ``raisin-pudding'',
where the cold HI clouds were the raisins while the neutral intercloud
medium was the pudding.  Now, we were told that most of the ISM was a
very hot dilute medium in which colder gas also existed (Cox and Smith
1976; Ostriker and McKee 1977).  The filling factor for this hot gas might
be very high (values of 90\% were suggested), which implies that the hot
gas will flow easily above the disk, creating a halo (it is now believed that the
filling factor is closer to 20\%).  Hot halo gas will not
persist in hydrostatic equilibrium due to radiative losses, so the evolution
of this gas was investigated, and the resulting weather pattern is referred to
as a galactic fountain (Shapiro and Field 1976; Bregman 1980).

Another great advance occurred beginning around 1980 with the launch of
the {\it International Ultraviolet Explorer\/} ({\it IUE\/}), which began the era of
ultraviolet absorption line astronomy of stars above the disk (and including
background AGNs; deBoer and Savage 1980).  These observations
revealed the presence of warm ionized gas species above the disk, with
temperatures somewhere in the range 10$^{{\rm 3.5}}$-10$^{{\rm 5}}$ K.  From these
observations, scales heights for the gas were deduced and we began to
understand the physical scale of the halo material.

The improvements in instrumentation in every waveband have led to
enormous advances in many aspects of this field.  A number of the leading
questions are on their way to being resolved, and I will discuss some of
these advances, many of which will be described in more detail by several
of the talks.  Finally, the relationship of this field of halo gas to other fields
of astrophysics has changed.  At one time, it was a well-contained subfield
that seemed to have little relationship to issues such as galaxy formation
and evolution.  Now, it is clear that many of the processes that we study
are closely related to the formation and evolution of galaxies, some of the
current ``big'' questions in astrophysics.

\section{The Size of the Gaseous Halo of the Milky Way}

One of the major issues in the field has been the location of the Galactic
halo gas.  For the neutral component, knowledge of the distance
determines the mass of the gas since the 21 cm measurement is a column
density.  The distance also determines the size of the clouds and of the
structures in which they reside, which extend across an entire quadrant of
the Galaxy in some cases.  Not only are the masses and sizes of the clouds
fundamental quantities, they place strong constraints on the models that
might produce such structures.

This has been one of the most fruitful research areas in the past several
years and we now have a good general idea of the distances of the HVC
and the intermediate velocity clouds (IVC).  The observational technique
has been to use multiple stars at different heights in the halo, along with
AGNs, which sample the full halo.  Against these objects, one can search
for absorption, generally using the optical absorption lines of Ca and Na
and a variety of UV lines.  For stars that are projected against a HVC, an
absorption line detection against a more distant star but with a non-detection 
for the closer star leads to a distance constraint, given certain
assumptions about the smoothness of the cloud.  There are many time-consuming 
steps in this process, such as identifying halo stars that have
sufficiently simple spectra and are bright enough for observations.  The
results have been worth the effort, and they show that the HVC lie at
distances of 5-20 kpc from the disk (Wakker et al. 2007, 2008; see
Wakker, this volume).  Related efforts have probed the IVC, which lie
closer to the disk, with typical distances of 0.5-4 kpc.  As described below,
the IVC are probably largely due to Galactic weather (a galactic fountain)
while the HVC (and especially the long coherent complexes) are more
likely due to accretion.

\begin{figure}
    \includegraphics[width=6cm, angle=0]{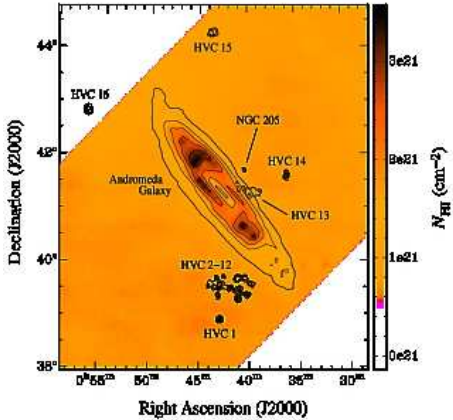}
    \caption{This grey scale image of the total HI column in M31 shows
    the extensive disk of the galaxy as well as gas surrounding the system 
    (Westmeier et al. 2007).  The gas surrounding M31 is probably the 
    counterpart to the HVC of the Milky Way.  The gas clouds around M31 lie
    within 50 kpc and generally closer.}
\end{figure}

If the Milky Way is not unique, HVCs should be present in other galaxies,
and the easiest cases to study are in the Local Group.  For M31, deep radio
synthesis maps show gas that can be HVC counterparts.  This gas is found
within 50 kpc of M31 but never beyond 50 kpc.  These clouds have
diameters of about 1 kpc and they masses of ${\sim}${}10$^{{\rm 5}}$ M$_{\odot}$ (Westmeier et al.
2007).  An important point is that there is no evidence for HI at hundreds
of kpc from the Milky Way or M31.  If there was neutral gas at such
distances, it would imply that there are large reservoirs of HI.  In this case,
the gas could be detected in other groups of galaxies, but searches for this
gas have only led to upper limits.

We have known about the scale heights of some ions for years, but the
picture has been filled out with observations of OVI (Savage et al. 2003),
the ion with the highest ionization potential among these ions.  Whereas
lower ionization state ions can be photoionized by starlight, the ionization
energy needed to produce OVI from OVI is 114 eV.  There is little
starlight above 114 eV as this lies well above the He II ionization energy
(54 eV), an important opacity in hot stars.  Therefore, this ion is
collisionally ionized and, in equilibrium, the peak of its ionization fraction
is at 10$^{{\rm 5.45}}$ K and the temperature range over which it is present is
relatively narrow (order of magnitude decrease from peak fractional
ionization occurs for T $<$ 10$^{{\rm 5.3 }}$ K, T $>$ 10$^{{\rm 5.7}}$ K).  
Gas near 10$^{{\rm 5.5}}$ K is near the
peak of the radiative cooling curve, so this gas should cool in less than 10$^{{\rm 7}}$
yr.  This gas either represents an intermediary cooling stage for gas that
began at a hotter temperature, or it is produced through turbulent mixing
or conductive interfaces with a hotter medium.  In either case, it implies
the presence of a yet hotter medium.

The radiative cooling time of this gas is less than the time for the gas to
free-fall onto the disk, so one would expect that cooler gas is produced,
and with a scale height similar to that of OVI.  This is indeed the case, as
similar scale heights are seen for a variety of species.  When making
comparisons between ions, there are a number of important details that
must be considered, such as the ionizing radiation field as a function of
height and the vertical distribution of cooling gas (which appears as an
effective source function in the fluid equations).

There is also evidence for hotter gas with a similar scale height in both the
Milky Way and in other galaxies.  In the Milky Way, the analysis of the
{\it Rosat All-Sky Survey \/}showed a diffuse emission component that followed
the Galactic disk (e.g., Pietz et al. 1998; Snowden et al. 2000).  It has a
scale height of about 4 kpc and a temperature near 10$^{{\rm 6}}$ K.  Thick disks of
hot gas are seen around external edge-on galaxies, with scale heights of 3-5 kpc.  
Taken at face value, this would appear to be the hotter phase that
radiatively cools, producing OVI, CIV, and the lower ionization ions that
are studied in absorption.  Around other galaxies, such as NGC 891, the
X-ray emitting gas is most prominent above the parts of the disk that have
active star formation, and this appears to be true for other galaxies. 
Furthermore, the intensity of these hot X-ray halos depends on the star
formation rate, suggesting that star formation, and their subsequent
supernovae are the driving elements for this gas (T\"{u}llmann et al. 2006). 
This would fit in with the predictions of galactic fountains.

\begin{figure}
    \includegraphics[width=10cm, angle=0]{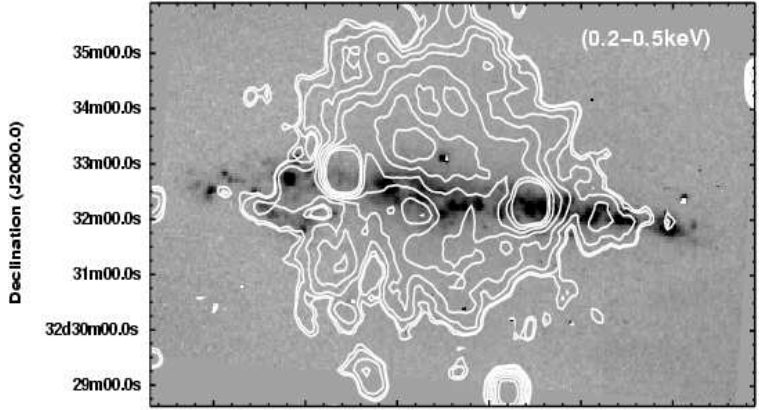}
    \caption{The H$\alpha$ image (black) denotes the active areas of the
    disk in the edge-on galaxy NGC 4631.  The white contours show the soft
    X-ray halo that surrounds the galaxy, extending several kpc from the disk
    (T\"{u}llmann et al. 2006).}
\end{figure}

A somewhat lower scale height is deduced from the H$\alpha$ survey (WHAM)
and the pulsar dispersion measurements (Reynolds, Haffner, Gaensler,
Berhkhuijsen, this meeting).  Depending on the assumptions that are made
for the filling fraction of electrons with height, values of 0.3 - 1 kpc are
obtained for the scale height, which might seem in conflict with the scale
heights for other halo components.  However, this is a single scale-height
fit, and for species that have strong disk components, the approach may be
too simplified.  The results do not exclude a more extended component,
and the electron density of the X-ray emitting component is estimated to
be 0.3-1$\times$10$^{{\rm -3}}$ cm$^{{\rm -3}}$, is an order of magnitude 
below their last data point, at
z = 1 kpc.  One topic where these dispersion measure values are crucial is
in constraining a more extended ionized halo.

Searching for an extended hot halo in emission has been challenging
because of current surface brightness limitations of X-ray astronomy. 
Early-type galaxies have hot halos that can be seen to 10-30 kpc
(Trinchieri, this meeting) and hot gas in starburst galaxies is seen to about
10-20 kpc, but for spiral galaxies, the greatest extent is only about 8 kpc,
in NGC 891.  A difficulty is that the emission measure decreases with the
square of the electron density and once the surface brightness drops to the
level of the X-ray foreground (due to the Milky Way and solar scattered X-rays) 
plus background (unresolved AGNs), further detection is impossible.

The situation is a bit more promising when making absorption line
observations of gas at the characteristic temperature of the Galactic
potential well, 1-3$\times$10$^{{\rm 6}}$ K.  The strongest lines in this temperature range
are the resonance lines of He-like and H-like oxygen, OVII and OVIII,
which occur at 574 eV and 654 eV.  The fractional equivalent widths of
absorption lines are inversely related to energy, and combined with the
limited effective collecting area of X-ray telescopes (55 cm$^{{\rm 2}}$ 
for {\it XMM-Newton\/}), these lines can be detected only against the continuum of the
brightest quasars.  We have determined the equivalent widths (or upper
limits) for the OVII resonance line for 17 extragalactic sitelines (Bregman
and Lloyd-Davies 2007).

We examined whether the equivalent width distribution was suggestive of
the Local Group geometry or of the Milky Way.  For the Local Group, one
would expect the greatest column densities to occur near the long axis and
in the general direction of M31 ({\it l\/} = 121$^{\circ}$), while the lowest columns
would occur in the anti-M31 direction.  In contrast, a Milky Way halo
would have the greatest columns in lines of sight that go across our
Galaxy.  The data are not consistent with a Local Group but are consistent
with a Milky Way halo (a hot medium must exist in the Local group,
possibly at a slightly higher temperature than can be sensed with OVII). 
The observations do not restrict the size of the halo, but a radius of 10-100
kpc is most likely; the implied gas mass is 10$^{{\rm 8}}$-10$^{{\rm 9}}$ M$_{\odot}$.  This size range
is consistent with the size of the HVC distributions around both the Milky
Way and M31.  There must be a dilute medium that fills the Local
Group and is most likely hot.  This medium is inferred from the 
apparent stripping of neutral gas from Local Group dwarfs that has left the 
ones within 250 kpc of the Milky Way or M31 as gas-poor (Blitz and Robishaw 2000; 
Putman, this conference).  Blitz and Robishaw (2000) infer a density of 
2.5$\times$10$^{-5}$ cm$^{-3}$ and a mass of 10$^{10}$ M$_{\odot}$ for this dilute medium.

A summary of the topics thus far is that there is excellent evidence for a
halo of gas with a characteristic thickness of 3-5 kpc.  This is a 
multi-temperature medium with a range of 10$^{{\rm 2}}$-10$^{{\rm 6}}$ K.  In addition, both neutral
and 10$^{{\rm 6}}$ K gas extends out to a radius of a few tens of kpc.  However, there
is no evidence for a large reservoir of neutral gas in the Local Group.

\section{Inflow Onto the Milky Way and Spiral Galaxies: Hot or Cold?}

The issue of accretion onto galaxies bears on several factors: evidence that
accretion has occurred or is occurring; an adequate reservoir of gas; and
whether the gas is accreting hot or cold.  Regarding accretion, there are
two modes, one in which the temperature of the accreted gas is less than
the characteristic temperature of the system.  For a spiral galaxy like the
Milky Way, the virial temperature is 2-3$\times$10$^{{\rm 6}}$ K and gas at this temperature
emits in the X-ray band.  Accretion of lower temperature gas cannot
naturally support itself by thermal pressure so it free-falls and radiates
through shocks.  In either case, the luminosity liberated is 

\hspace*{1.28cm}L = 4$\times$10$^{{\rm 40}}$ (Mdot/1 M$_{\odot}$ yr$^{{\rm -1}}$) (T/3$\times$10$^{{\rm 6}}$ K) erg/sec\\
In considering the hot accretion mode, we note that the Milky Way has L$_{{\rm X}}$
$\approx$10$^{{\rm 39.3}}$ erg s$^{{\rm -1}}$, and the X-ray luminosity from other spirals is about 10$^{{\rm 39.5}}$
erg s$^{{\rm -1}}$, most of which appears to be related to star formation in the disk.  If
this X-ray emission is due to accretion and not a galactic fountain, then the
hot accretion rate is no more than 0.2 M$_{\odot}$ yr$^{{\rm -1}}$. However, T\"{u}llmann et al.
(2006) has shown that this X-ray emission is more likely galactic in origin,
in which case, the hot accretion rate is probably less than 0.1 M$_{\odot}$ yr$^{{\rm -1}}$.

In contrast, there is considerable evidence for cold accretion onto spiral
galaxies at a rate of about 0.2 M$_{\odot}$ yr$^{{\rm -1}}$. The HVC, including the
Magellanic Stream is neutral material falling onto the Milky Way.  In a
few cases, we see evidence for a collision between the HI and the Milky
Way disk, as in the case of the Smith Cloud (Lockman et al. 2008) and the
leading arm of the Magellanic Stream (McClure-Griffiths et al. 2008). 
More evidence for cold accretion comes from the distortions in the HI
disks of spiral galaxies.  Without accretion events, the disks of spirals
would be approximately circularly symmetric, but a significant fraction
show distortions that are best understood as the accretion of cold material
onto the outer parts of the galaxy (review by Sancisi et al. 2008).  The
inferred accretion rate is 0.2 M$_{\odot}$ yr$^{{\rm -1}}$, the same as the value deduced for
the Milky Way.

There are two arguments that are often made for an accretion rate that is
about an order of magnitude larger, about 1-2 M$_{\odot}$ yr$^{{\rm -1}}$. The first is the
desire to resolve the G-dwarf problem, where there are too few G dwarfs
with low metallicity (the metallicity ``floor'' is about 0.2 Z$_{\odot}$).  A popular
solution to the G dwarf problem is to have a steady inflow of low
metallicity material onto the disk, at about a rate of 1-2 M$_{\odot}$ yr$^{{\rm -1}}$.  This is
not the only possible resolution to this problem and alternative
explanations seem viable (Binney and Merrifield 1998).  A natural
solution to the G-dwarf problem is that the gas that formed the stellar disk
was pre-enriched.

The other problem that accretion would resolve is the gas consumption
problem.  The problem for the whole galaxy is that, at the current star
formation rate, most of the gas will be consumed in ${\sim}${}10$^{{\rm 9.5}}$ Gyr.  This does
not seem to be a serious a problem because the volume-averaged star
formation rate of the universe is decreasing sharply in time rather than
being constant.  However, for the inner part of a galaxy, the gas depletion
timescale can be several times shorter (Blitz, private communication). 
Unless there is gas replenishment, star formation will diminish on a
timescale less than 1 Gyr.  In principle, gas replenishment could take the
form of gas transfer from the outer to the inner part of a spiral galaxy,
although models do not yet predict a smooth mass transfer rate ${\sim}${}1 M$_{\odot}$
yr$^{{\rm -1}}$.  Such mass transfer might occur in large events, such as when
Magellanic Stream material reaches the disk.  This could lead to periods of
significant mass transfer and a variable star formation rate.  In the region
near the Sun, the star formation rate has varied significantly (Rocha-Pinto
et al. 2000), and in following a period of enhanced star formation 2 Gyr
ago, the star formation rate dropped by a factor of four in a fraction of a
Gyr.  If radial mass transfer is ineffective, then accretion would need to
occur onto the star-forming parts of spiral galaxies.   The accreted gas
would need to have just the right angular momentum distribution for this
to work.  In this area, I would like to see more theoretical work regarding
the effectiveness of radial mass transfer and whether an accretion rate
greater than the observed value of 0.2 M$_{\odot}$ yr$^{{\rm -1}}$ is really required.  A
radial flow rate through the disk of 0.2 M$_{\odot}$ yr$^{{\rm -1}}$ would require a flow
velocity of no more than 1 km s$^{{\rm -1}}$, which is below the limitations of such
measurement, about 5-10 km s$^{{\rm -1}}$ (Wong et al. 2004).

\begin{figure}
    \includegraphics[width=7cm, angle=-90]{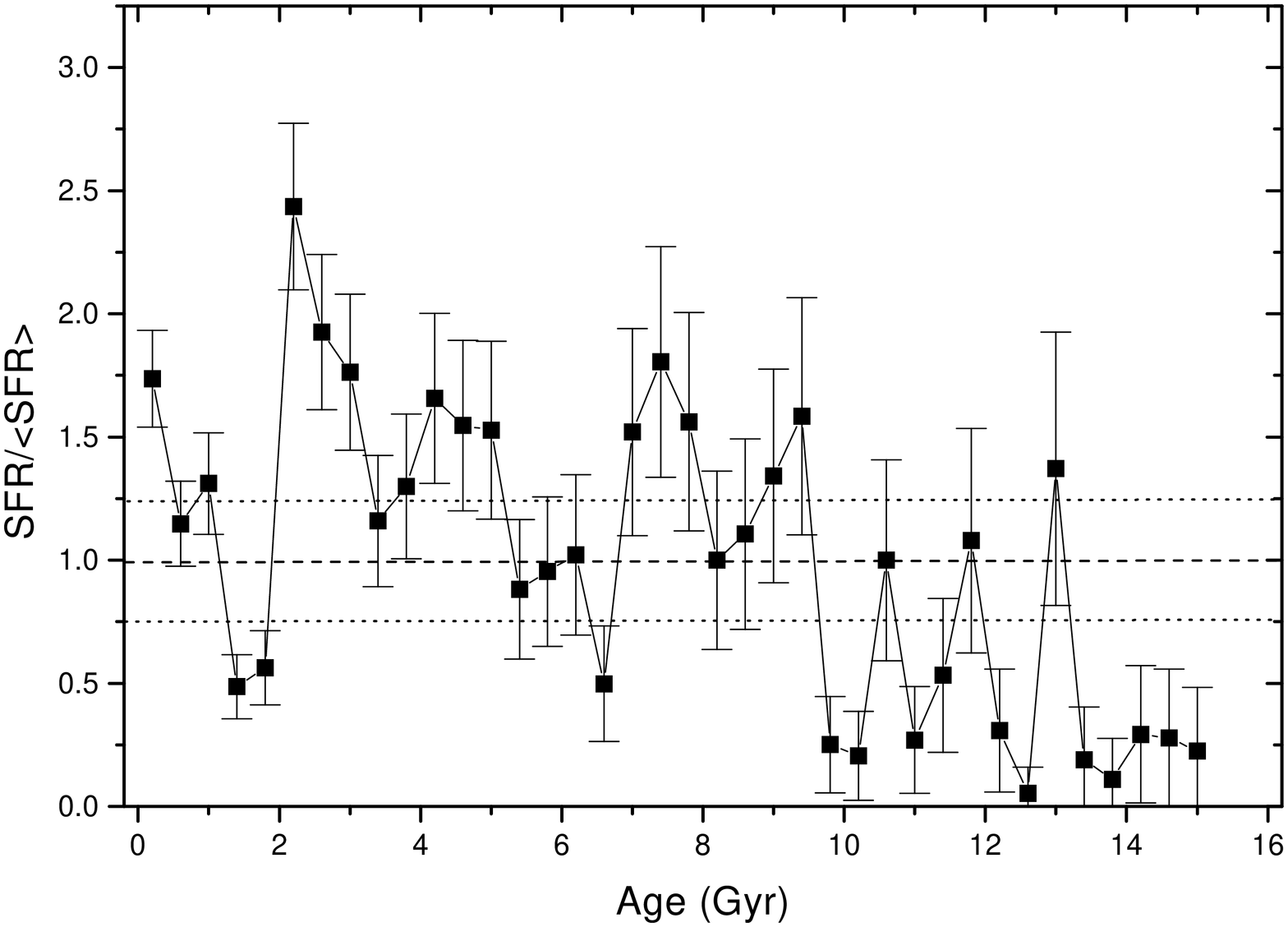}
    \caption{The historical star formation rate in region of the Sun, based
    on chromospheric ages.  The dotted lines show the 2$\sigma$ bounds for
    a constant star formation rate model (Rocha-Pinto et al. 2000).  
    Note some rapid changes in the rise and fall of the star formation rate, 
    such as the decrease by a factor of four that occurred 2 Gyr ago.}
\end{figure}

A related, and rather serious problem is that of the reservoir of the
accreting gas.  X-ray observations of galaxy groups and of spiral galaxies
show that they are relatively poor in hot gas.  As discussed above, the
cooling rate is probably insufficient to supply accretion from the cooling
flow mechanism.  An alternative is that the gas is neutral, but HI surveys
do not find a reservoir of 10$^{{\rm 9}}$-10$^{{\rm 10}}$ M$_{\odot}$ of cold gas within a virial radius
of galaxies (or a larger amount in galaxy groups).  If such a reservoir of
gas exists, it is either molecular or it is warm ionized material.  A reservoir
of molecular gas seems less likely because the gas that disturbs the
outskirts of spiral galaxies and the HVC are primarily neutral atomic gas. 
The remaining possibility is a reservoir of warm ionized gas (10$^{{\rm 3}}$-10$^{{\rm 4}}$ K),
which would be a very low surface brightness emitter at the low pressures
of the halo.  This warm ionized material would be in pressure equilibrium
with a hot ambient medium.  If there is an extended hot halo around a
spiral galaxy, there would be a negative pressure gradient until the material
merged with the group medium.  Thus, as the gas falls toward the galaxy,
its density would increase and eventually recombine, as the recombination
rate exceeds the photoionization rate.

Although it may be difficult to detect this material in emission, there is
some evidence for it in absorption.  Wakker (this volume) has examined
the distance of low-redshift absorption systems from the nearest galaxy
and finds that about half of these systems lie within 400 kpc of the host
system.  There is a lot of mass in such low redshift absorption systems: the
ones within 400 kpc of the galaxies contain about three times the amount
of baryons as are in the galaxies.  This seems to be an adequate supply of
gas, provided that it can fall in at the observed rate.  To conclude, the most
likely form for the reservoir of gas that falls onto spiral galaxies is warm
ionized material.

\section{Cold Flows, A New Paradigm?}

A shift is occurring in the field of galaxy formation that may bear directly
on some of the accretion issues discussed above.  For massive galaxies, the
formation mechanism has resembled cooling flows, whereby gas falls into
the dark matter potential, heats up to the characteristic temperature of the
potential well in a standoff shock, and then radiatively cools to form the
galaxy.  The problem with this attractive theory is that it leads to relatively
slow galaxy formation and it is unable to produce the luminous galaxies
that are observed at early cosmological times.  A solution to this problem
was offered by Dekel et al. (2009), who argued that there are filaments of
low entropy gas that, because of lack of buoyancy,  would flow into the
galaxy quickly.  These cool filaments penetrate the standoff shocks in the
hot gas, feeding the galaxy with cooler gas.  Such cold flows are now
being identified in high resolution simulations, although it may be a while
before their importance is fully understood.
The reason that I bring up the topic of cold flows is that the accretion onto
galaxies today may be a very modest version of this phenomenon.  At early
epochs, cold flows would need to occur at a rate of ${\sim}${}10$^{{\rm 1}}$-10$^{{\rm 2}}$ M$_{\odot}$ yr$^{{\rm -1}}$ to
produce the galaxies seen at high redshift.  If the regions surrounding these
galaxies are not greatly disturbed (as would occur in a galaxy cluster), then
it seems feasible that a more modest amount of material, initially near or
beyond the virial radius, is falling in today.  Understanding cold flows in
greater depth may be quite important for interpreting the observations of
galaxies today.  
\begin{figure}
    \includegraphics[width=7cm, angle=0]{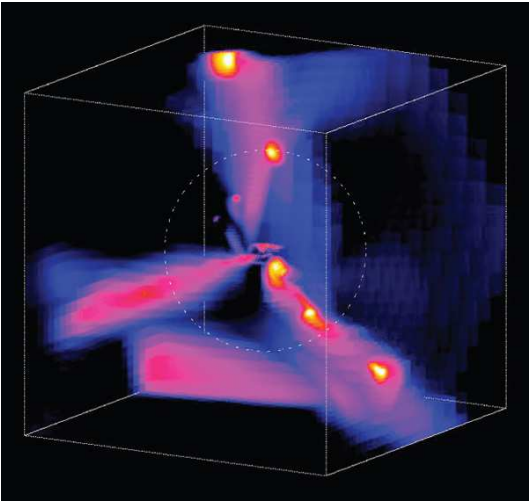}
    \caption{In this simulation cube of 320 kpc, low-entropy flows (cold flows)
    occur in three filaments.  The intensity reflects the inflow rate onto
    the galaxy (Dekel et al. 2009).}
\end{figure}

\begin{figure}
    \includegraphics[width=10cm, angle=0]{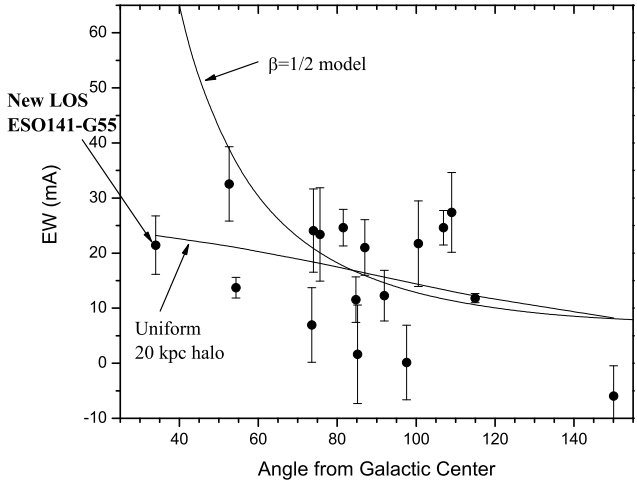}
    \caption{The distribution of OVII absorption lines as a function of angle
    from the Galactic Center.  The OVII line is produced by 10$^6$ K gas in a 
    halo around the Milky Way.  The new data indicates that a dilute extended halo
    is preferred to a halo where there is significant gas bound to the bulge, as
    is seen in elliptical galaxies (Lloyd-Davies and Bregman, in preparation).}
\end{figure}

\section{Galactic Winds}

In addition to the infall of gas onto galaxies, there may be outflows as
well.  It seems that many astronomical objects both accrete and eject gas
and spiral galaxies may not be very different.  There are two likely
mechanisms for the outflow of gas: cosmic rays; and thermally driven
winds.  The thermally driven winds occur because supernovae, and
possibly AGN heating, raises the gas temperature above the escape
temperature.  This is the likely mechanism that keeps many early-type
galaxies relatively free of gas and X-ray poor (especially the elliptical
systems with L $<$ L*; Trinchieri, this meeting).  It would not be too
surprising if it occurred in the bulge of the Milky Way and the energetics
are consistent with this picture.

Hot gas is present in the direction of the bulge as seen in all-sky X-ray
surveys and in X-ray shadowing studies.  If the gas were bound to the
bulge, it would follow a radial density distribution similar to that seen in
X-ray luminous elliptical galaxies, {\it S$_{{\rm x}}$ $\propto$ (1+(r/r$_{{\rm c}}$)$^{{\rm 2}}$)$^{{\rm -3}}$\/}$^{{\rm \beta}{{\it +1/2\/}}}$, where $\beta$ = 0.5-0.7,
typically.  This leads to a density distribution of hot gas of {\it n\/} $\propto$ {\it (1+(r/r$_{{\rm c}}$)$^{{\rm 2}}$)$^{{\rm -3}}$\/}$^{{\rm \beta}{{\it /2\/}}}$.  This implies that the density and column density of hot gas should rise
rapidly into the bulge of the Milky Way, but no such rise is observed.  This
is consistent with an outflow of hot gas from the bulge and the outflow
rate would be 0.1 M$_{\odot}$ yr$^{{\rm -1}}$ if all of the mass shed from the stars partakes in the wind.

A cosmic-ray driven wind, extending beyond the bulge and into the disk is
discussed at this meeting by Breitschwerdt and Evrett.  This seems
reasonable, but one of the frustrations is that we do not have sufficiently
good knowledge of cosmic rays and magnetic fields to accurately predict
the mass flux of such a wind (although the new calculations presented here
are quite promising).

The winds from the Milky Way or even a low redshift starburst galaxy
pales by comparison with some high redshift galaxies, which may be
hundreds of times greater.

\section{The Missing Baryons in Galaxies}

There are two missing baryon problems, one for the local universe and
another for individual galaxies.  To appreciate the first problem, we
consider the dark matter and the baryonic matter, where we can form the
ratio known as the baryon fraction.  This fraction is given from WMAP
observations, but also from Big Bang Nucleosynthesis calculations, and
the baryon fraction is about 17\%, or ${\Omega}_{baryon}$ = 0.046 
and ${\Omega}_M$ = 0.27.  It had
been assumed that the baryons all went into galaxies and if we took a
census today, we would obtain the same ratio.  However, when one
examines the amounts of baryons in different components, one finds very
little in the form of galaxies.  Galaxies at low redshift constitute only
about 4\% of ${\Omega}_{baryon}$, with another 3\% in hot gas in X-ray bright clusters and
groups of galaxies (Fukugita and Peebles 2004).  About 30\% of ${\Omega}_{baryon}$
resides in warm gas responsible for Ly$\alpha$ absorption line systems and
another 5-10\% in OVI - bearing gas (about 10$^{{\rm 5.5}}$ K; Danforth \& Shull
2005)).  That leaves about 50\% unaccounted for and this is the ``missing''
baryon problem.  These baryons are not really missing but are simply
difficult to detect.  Theory suggests that this material is hot 
(10$^5$-10$^7$ K)
and dilute, with an overdensity of 10-100 and it has not collapsed into
virialized structures (Cen and Ostriker 2006).  This gas is expected to lie
in the large unvirialized filaments that connect galaxy clusters and galaxy
groups.  If this is the correct model, it will be detected when X-ray
absorption spectroscopy improves by an order of magnitude or more.

There is a second missing baryon problem that is much more local, being
evident around individual galaxies.  If you add together the stars and gas in
the Milky Way and divide by the gravitational mass, you only account for
20-30\% of ${\Omega}_{baryon}$.  This is not an isolated incident.  It applies to every
galaxy.  Another nearby galaxy, M33 only has 0.1 ${\Omega}_{baryon}$ (Corbelli 2003),
and gravitational lensing studies also show that only 10-30\% of baryons lie
in moderate and large galaxies (e.g., Hoekstra et al. 2005, McGaugh 2007). 
The situation is even worse in the dwarf galaxies, many of which possess
less than 0.1 ${\Omega}_{baryon}$.  {\it Relative to their dark matter content, galaxies are baryon-poor\/}.

\begin{figure}
    \includegraphics[width=10cm, angle=0]{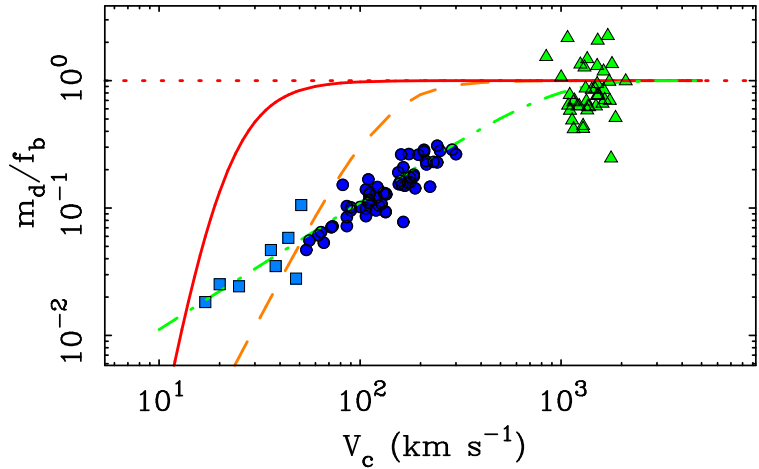}
    \caption{The ratio of the baryon fraction to the cosmic value of the baryon
    fraction as a function of circular velocity (McGaugh 2007).  The Rich clusters 
    of galaxies contain the cosmic value of baryons, but the shallower potential
    well systems, galaxies, are missing most of their baryons.  The loss of
    baryons becomes extreme for the dwarf galaxies.}
\end{figure}

One might wonder whether any present day structure has the cosmological
baryon-to-dark matter ratio.  Deep potential wells, such as clusters of
galaxies have nearly the cosmological value of this ratio.  For potential
wells deeper than about 1.5 keV (2$\times$10$^{{\rm 7}}$ K or 500 km/s rotational velocity),
which are rich galaxy groups, the baryons are not missing.

There are two likely scenarios for the baryons absent in galaxies.  Either
the gas fell into the galaxy but was heated by supernovae and AGNs,
leading to a galactic superwind.  There is evidence for galactic superwinds
at high redshift (z = 2-3) with outflow velocities of 300-1000 km/sec and
mass loss rates of 100 M$_{\odot}$ yr$^{{\rm -1}}$.  If this were to persist for 1 Gyr, it could
account for the amount of baryons missing from galaxies.  Unfortunately,
the duration of such superwinds is unknown and even their mass fluxes are
not accurately known.  This superwind would pollute the surrounding
medium with metals and it would heat the gas.

Another possibility is that the gas is heated before falling into the galaxy. 
If enough supernovae occur prior to the primary collapse of the galaxy
components, the gas may have enough entropy to resist the infall.  The
energy needed to prevent the infall of gas is quite considerable, but it is
about a factor of two less than the energy needed to drive it out as a
galactic wind.  If this gas resided within the virial radius of a galaxy like
the Milky Way, it would have a very massive halo, with a mass of about 1-2$\times$10$^{{\rm 11}}$ M$_{\odot}$ within a radius of 200 kpc, the approximate virial radius. 
Such hot halos are not seen around galaxies, so the material must have
been pushed well beyond the virial radius.  Since the mass a dilute Local
Group medium is estimated to be only 10$^{10}$ M$_{\odot}$ (Blitz and Robishaw 2000),
it is likely that this ejected gas lies beyond the virial radius of the Local Group.

These important heating or feedback processes are generally not yet
included in galaxy formation models, which typically predict that galaxies
form with the cosmological baryon fraction.  It is only when these
processes are properly accounted for, along with the development of cold
flows, that we can understand galaxies and the accretion or outflow
processes that may be occurring today.

The problems and issues that I outlined above do not constitute an exhaustive
list of topics represented by this meeting, or even a list of the most interesting
areas.  They are a personal selection of some of the issues that I would like
to know the answers to.  Judging from the posters and talks in this meeting,
we will learn a great deal about the progress in these areas, as well as 
many others.

I would like to thank the local organizing committee for their hospitality
and efforts in organizing a exciting meeting.  Also, I gratefully
acknowledge financial support from NASA.


\end{document}